\documentclass{article}

\usepackage{PRIMEarxiv}

\usepackage[utf8]{inputenc} % allow utf-8 input
\usepackage[T1]{fontenc}    % use 8-bit T1 fonts
\usepackage{hyperref}       % hyperlinks
\usepackage{url}            % simple URL typesetting
\usepackage{booktabs}       % professional-quality tables
\usepackage{amsmath}
\usepackage{amsthm}
\usepackage{amssymb}
\usepackage{amsfonts}
\usepackage{mathtools}
\usepackage{amsfonts}       % blackboard math symbols
\usepackage{nicefrac}       % compact symbols for 1/2, etc.
\usepackage{microtype}      % microtypography
\usepackage{fancyhdr}       % header
\usepackage{graphicx}       % graphics
\graphicspath{{media/}}     % organize your images and other figures under
\usepackage{times}
\usepackage{subfig}
\usepackage{xcolor}
\usepackage{array}
\usepackage{cite}
\usepackage{multirow}
\usepackage{adjustbox}
\usepackage[scr=dutchcal]{mathalfa}
\usepackage[font=small,labelfont=bf]{caption}

\DeclareMathOperator{\diag}{diag}

\DeclareMathOperator{\InFlow}{in}
\DeclareMathOperator{\OutFlow}{out}
\DeclareMathOperator{\pump}{pump}
\DeclareMathOperator{\col}{col}
\DeclareMathOperator{\row}{row}

\newtheorem{theorem}{Theorem}
\newtheorem{lemma}{Lemma}

\newtheorem{definition}{Definition}

\newtheorem{remark}{Remark}

\usepackage{tikz-network}
\tikzset{
    block/.style = {draw, fill=white, rectangle, minimum height=.75cm, minimum width=1.25cm},
    tmp/.style  = {coordinate}, 
    sum/.style= {draw, fill=white, circle},
    input/.style = {coordinate},
    output/.style= {coordinate},
    pinstyle/.style = {pin edge={to-,thin,black}
    }
}

\title{Distributed Estimation with Decentralized Control for Quadruple-Tank Process
\thanks{Empty.} 
}

\author{
  Moh. Kamalul Wafi, Bambang L. Widjiantoro \\
  Department of Engineering Physics \\
  Institut Teknologi Sepuluh Nopember \\
  Surabaya, Indonesia \\
  \texttt{\{kamalul.wafi\}@its.ac.id} \\
}

\begin{document}
\maketitle

\begin{abstract}
This paper presents a unified modeling, control, and estimation framework for the quadruple-tank process—a benchmark multivariable system that exhibits either minimum-phase or non-minimum-phase behavior depending on valve flow ratios. A decentralized PI control strategy is employed to regulate water levels, while a distributed state estimation scheme is developed using local Luenberger observers and inter-agent communication. Each observer uses only local output measurements and exchanges information with neighboring nodes over a strongly connected communication graph. To address the limitations of partial observability, the observer design incorporates an observability decomposition and consensus-based coupling that ensures convergence to the true system state. Simulation results validate the effectiveness of the proposed framework, demonstrating accurate state reconstruction and stable closed-loop performance under both minimum-phase and non-minimum-phase configurations. These results highlight the potential of combining decentralized control with distributed estimation for scalable, networked control of complex multivariable systems.
\end{abstract}

\allowdisplaybreaks

\section{Introduction}
The analysis and control of complex multivariable dynamical systems have garnered significant attention in modern control theory, particularly in benchmark processes such as the quadruple-tank system \cite{R1,R2}. This laboratory-scale setup offers a rich platform to study the interplay between nonlinear dynamics, interaction effects, and performance limitations in multivariable settings. Notably, the quadruple-tank system can be configured to exhibit either minimum-phase or non-minimum-phase behavior, depending on the ratio of valve openings. This structural property makes it ideal for exploring control limitations and algorithm design.

A central challenge in such systems lies in the interconnection between subsystems: the state of one tank directly affects the others, thereby introducing coupling that complicates control and estimation. Ensuring closed-loop stability in the presence of this coupling often requires maintaining all system poles in the left-half of the complex plane. Various control strategies have been proposed to address these challenges, including sliding mode control \cite{R6}, robust control \cite{R7}, and more recently, model predictive control \cite{R8}. In parallel, system identification approaches—particularly batch estimation techniques—have been used to derive accurate models for control synthesis \cite{R3}. These methods provide insights into the achievable performance, especially under non-minimum-phase conditions \cite{R4}, further elaborated in \cite{R5}.

Decentralized control, where controllers operate with local information while coordinating through a predefined structure, has also proven effective in this context \cite{R9,R10}. In the case of the quadruple-tank system, this approach allows independent PI controllers for each input-output channel, leveraging structural knowledge to linearize the dynamics around a desired operating point \cite{R2}. When the system is configured in the non-minimum-phase setting, control design becomes significantly more demanding, requiring careful gain tuning to avoid instability or degraded performance.

Beyond control, the need for state estimation is equally critical, especially in distributed settings where each node only has access to partial output measurements. Filtering techniques such as the Kalman filter \cite{R11,R14,R15} have been adapted to this distributed framework, leading to innovations in distributed estimation and fusion. Early works applied centralized estimation techniques like track fusion using cross-covariance \cite{R16}, followed by extensions incorporating maximum likelihood estimation \cite{R17} and consensus-based filtering \cite{R18,R19}. These efforts laid the foundation for distributed state estimation where each node—typically associated with an individual output—employs a local estimator and exchanges information with its neighbors.

Distributed Luenberger observer designs have since been proposed for linear systems under various topologies \cite{R12,R13,R20,R21,R22}. These methods aim to reconstruct the global system state through local measurements and neighbor communication, often leveraging graph-theoretic insights and observability decompositions. The design in \cite{R12}, in particular, is extended in \cite{R23} to handle discrete-time dynamics, drawing inspiration from the observability structures described in \cite{R24}. The key conditions under which distributed observers achieve omniscience—where all state estimates converge to the true state—are discussed in \cite{R24}. In essence, if a node's local output is insufficient to observe the system independently, information from neighboring nodes must be incorporated through a communication graph satisfying detectability conditions \cite{R5}.

This paper builds upon these foundations by combining a decentralized control structure with a distributed estimation framework tailored to the quadruple-tank system. First, a mathematical model is presented that captures both minimum-phase and non-minimum-phase configurations. A decentralized PI controller is designed based on this model, with performance evaluated under both scenarios. Next, a distributed observer based on local Luenberger structures is proposed, relying only on local outputs and neighbor communications. Finally, numerical simulations validate the approach and demonstrate the observer's effectiveness across varying conditions. The paper concludes with a discussion of future directions, including extensions to fault-tolerant estimation and control.

\subsection*{Notation}

For a matrix $M$, its transpose is denoted by $M^\top$ and its inverse by $M^{-1}$. The symmetric part of $M$ is defined as $\operatorname{Sym}(M) := M + M^\top$, and $\operatorname{rank}(M)$ denotes its rank. The identity matrix of dimension $N$ is written as $I_N$, and the vector of all ones of size $N \times 1$ is denoted by $\mathbf{1}_N$.

For a symmetric matrix $P$, the notation $P > 0$ ($P < 0$) implies that $P$ is positive (negative) definite. Given a set of matrices $\{A_1, A_2, \dots, A_N\}$, we write $\diag\{A_1, A_2, \dots, A_N\}$ to indicate a block diagonal matrix with $A_i$ along the diagonal, and use $\col(A_1, A_2, \dots, A_N)$ to denote the block column formed by vertically stacking these matrices. Similarly, $\row(A_1, A_2, \dots, A_N)$ refers to the block row matrix $[A_1^\top \; A_2^\top \; \cdots \; A_N^\top]^\top$. The Kronecker product of matrices $M_1$ and $M_2$ is expressed as $M_1 \otimes M_2$.

Let $A: \mathcal{X} \to \mathcal{Y}$ be a linear map between finite-dimensional vector spaces. The kernel of $A$ is given by $\ker A := \{x \in \mathcal{X} \mid Ax = 0\}$, and the image is $\operatorname{im} A := \{Ax \in \mathcal{Y} \mid x \in \mathcal{X}\}$. For a subspace $\mathcal{V} \subset \mathcal{X}$, the orthogonal complement is denoted $\mathcal{V}^\perp$.

We consider a graph $\mathcal{G} = (\mathcal{N}, \mathcal{E}, \mathcal{A})$, where $\mathcal{N} = \{1, 2, \dots, N\}$ is a set of nodes, $\mathcal{E} \subset \mathcal{N} \times \mathcal{N}$ is a set of directed edges, and $\mathcal{A} = [a_{ij}] \in \mathbb{R}^{N \times N}$ is the adjacency matrix, where $a_{ij} > 0$ if $(i, j) \in \mathcal{E}$ and $a_{ij} = 0$ otherwise. An edge $(i,j) \in \mathcal{E}$ indicates that information flows from node $i$ to node $j$. The graph is undirected if $(i,j) \in \mathcal{E}$ implies $(j,i) \in \mathcal{E}$ for all $i, j \in \mathcal{N}$. A graph is connected if, for any distinct nodes $i, j \in \mathcal{N}$, there exists a path from $i$ to $j$.

The Laplacian matrix $\mathcal{L} = [\ell_{ij}] \in \mathbb{R}^{N \times N}$ is defined by $\mathcal{L} := \mathcal{D} - \mathcal{A}$, where $\mathcal{D}$ is the diagonal matrix of node degrees given by $d_i := \sum_{j \in \mathcal{N}} a_{ij}$. By construction, $\mathcal{L}$ has a zero eigenvalue with right eigenvector $\mathbf{1}_N$, i.e., $\mathcal{L} \mathbf{1}_N = 0$, and all remaining eigenvalues lie in the right-half complex plane. For undirected graphs, $\mathcal{L}$ is symmetric and positive semidefinite. The eigenvalues of $\mathcal{L}$ are nonnegative and ordered as $0 = \lambda_1 < \lambda_2 \leq \dots \leq \lambda_N$, where $\lambda_2 > 0$ if and only if the graph is connected.

For a vector $x \in \mathbb{R}^n$, the Euclidean norm is denoted by $\|x\|$, and for a matrix $P \in \mathbb{R}^{n \times n}$, the induced 2-norm is denoted by $\|P\|$. Given a set $\Omega \subset \mathbb{R}^n$, the complement and closure are denoted $\Omega^c$ and $\overline{\Omega}$, respectively. For a positive scalar $\epsilon > 0$, the $\epsilon$-neighborhood of $\Omega$ is denoted $\mathcal{B}_\epsilon(\Omega) := \{x \in \mathbb{R}^n \mid \exists y \in \Omega, \|x - y\| < \epsilon\}$.

\section{System Modeling and Dynamic Behavior}\label{Sec:2}
\begin{figure}
    \centering
    \includegraphics[width=0.45\linewidth]{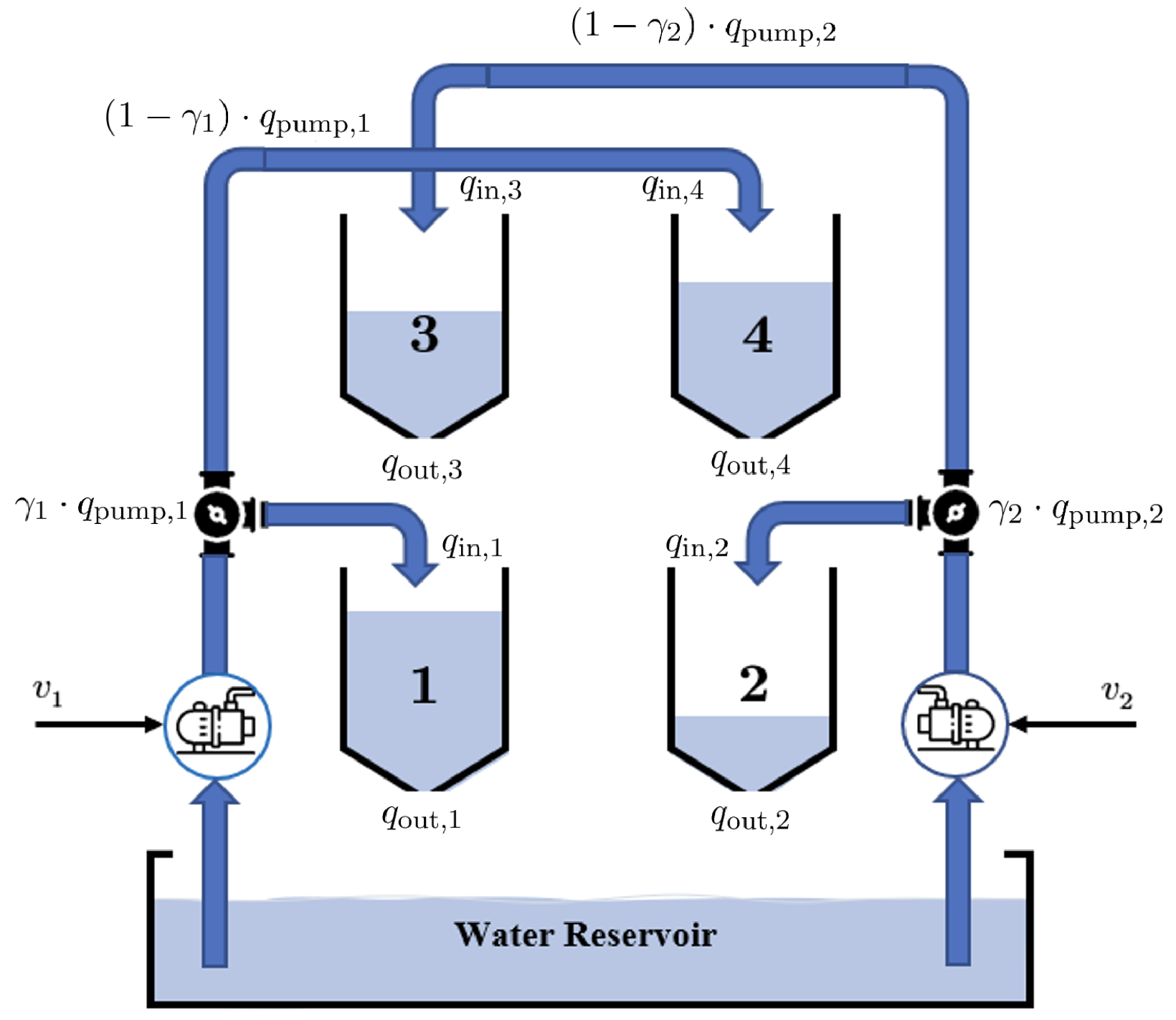}
    \caption{The design of quadruple-tank process}
    \label{Fig1}
\end{figure}

The quadruple-tank system comprises four interconnected tanks, indexed by $i = 1,\dots,4$, and is actuated by two pumps, indexed by $p = 1,2$, as illustrated in Fig.~\ref{Fig1}.
This setup forms a classic benchmark for multivariable control due to its non-minimum phase behavior and strong cross-coupling. It is modeled as a multivariable-input, multivariable-output (MIMO) process with two control inputs and two measured outputs. The control inputs are the voltages applied to the pumps, denoted as \( v_1 \) and \( v_2 \), while the outputs correspond to the water levels in tanks 1 and 2, measured as \( h_1 \) and \( h_2 \), respectively. Measurement sensors with gain $k_c$ are installed in the lower two tanks (1 and 2), and the primary objective is to regulate their liquid levels $(h_1, h_2)$ to desired setpoints. 

Each pump delivers water through a three-way valve, which splits the flow into two branches directed to diagonally opposite tanks. The distribution of the water from pump $p = 1,2$, is governed by the valve ratios $\gamma_p \in [0,1]$, which define the flow split between the tanks. Let the voltage applied to pump $p$ be $v_p$, and the resulting flow rate be given by:
\begin{equation*}
    q_{\pump, p} = k_p v_p,
\end{equation*}
where \( k_p \) is the pump gain (a constant), and \( v_p \) is the input voltage.

The distribution of flow to the tanks is regulated by valve settings \( \gamma_1, \gamma_2 \in [0,1] \), representing the fraction of the flow directed to specific tanks:
\begin{itemize}
    \item Pump 1 delivers \( \gamma_1 k_1 v_1 \) to tank 1 and \( (1 - \gamma_1) k_1 v_1 \) to tank 4.
    \item Pump 2 delivers \( \gamma_2 k_2 v_2 \) to tank 2 and \( (1 - \gamma_2) k_2 v_2 \) to tank 3.
\end{itemize}

To derive the system dynamics, we apply the principle of mass conservation. For each tank $i$, the rate of change of fluid mass equals the net difference between the inflow and outflow mass rates:
\begin{equation}\label{Eq:Dyn:1}
    \frac{d m_{T_i}}{dt} = \sum\nolimits_{j\in\mathcal{N}_i} m_{\InFlow,j} - m_{\OutFlow,i},
\end{equation}
where $m_{T_i}$ [kg] denotes the mass of fluid in tank $i$, and $m_{\InFlow,j}$ [kg/s] and $m_{\OutFlow,i}$ [kg/s] represent the inlet and outlet mass flow rates, respectively. Here, $\mathcal{N}_i$ denotes the set of upstream sources supplying inflow to tank $i$, indexed by $j$.

Assuming the fluid is incompressible with constant density $\rho = \rho_i$ [kg/m³] for all $i$, and using the relations $m_T = \rho V_i = \rho A_i h_i$ and $m_{\text{in/out}} = \rho q_{\text{in/out}}$, where $A_i$ is the cross-sectional area of tank $i$, $h_i$ is the liquid height, and $q_{\text{in/out}}$ are the volumetric flow rates [m³/s], we obtain:
\begin{equation}\label{Eq:Dyn:2}
    \frac{d}{dt} (\rho A_i h_i) = \sum\nolimits_{j\in\mathcal{N}_i}\rho q_{\InFlow,j} - \rho q_{\OutFlow,i}.
\end{equation}
Dividing both sides by $\rho$, the equation simplifies to:
\begin{equation}\label{Eq:Dyn:3}
    A_i \frac{dh_i}{dt} = \sum\nolimits_{j\in\mathcal{N}_i} q_{\InFlow,j} - q_{\OutFlow,i}.
\end{equation}
The inlet flow rates to each tank are defined by the pump-valve configuration as:
\begin{align}\label{Eq:Dyn:4} 
    \begin{aligned}
    q_{\InFlow,1} \coloneqq \gamma_1q_{\pump,1} &= \gamma_1 k_1 v_1, \quad q_{\InFlow,4} \coloneqq (1-\gamma_1)q_{\pump,1}= (1 - \gamma_1) k_1 v_1, \\
    q_{\InFlow,2} \coloneqq \gamma_2q_{\pump,2} &= \gamma_2 k_2 v_2, \quad q_{\InFlow,3} \coloneqq (1-\gamma_2)q_{\pump,2}= (1 - \gamma_2) k_2 v_2. \end{aligned}
\end{align}
The outlet flow at the bottom of each tank follows Torricelli's law, derived from Bernoulli’s principle, which relates pressure head to exit velocity:
\begin{equation}\label{Eq:Dyn:5}
    q_{\OutFlow,i} = a_i \sqrt{2 g h_i},
\end{equation}
where \( a_i \) is the cross-sectional area of the outlet orifice [m²], and $g$ is the gravitational constant [m/s²].

Note that the set $\mathcal{N}_i$ is introduced to account for the fact that the dynamics of the lower tanks are influenced not only by the direct inflow but also by return flows from the upper tanks. Specifically, tanks 1 and 2 receive fluid not only from their respective pumps but also from the outlets of tanks 3 and 4, respectively. These interactions are governed by the structure of the interconnection and the diagonal influence induced by the pump configuration. Thus, the full dynamic equations of the tank levels are:
\begin{align}\label{Eq:Dyn:6}
    \begin{aligned}
    A_1 \frac{dh_1}{dt} &= q_{\InFlow,1} + q_{\OutFlow,3} - q_{\OutFlow,1}, \\
    A_2 \frac{dh_2}{dt} &= q_{\InFlow,2} + q_{\OutFlow,4} - q_{\OutFlow,2}, \\
    A_3 \frac{dh_3}{dt} &= q_{\InFlow,3} - q_{\OutFlow,3}, \\
    A_4 \frac{dh_4}{dt} &= q_{\InFlow,4} - q_{\OutFlow,4}.
    \end{aligned}
\end{align}
Substituting the inflow expressions from \eqref{Eq:Dyn:4} and the outflow expressions from \eqref{Eq:Dyn:5} into \eqref{Eq:Dyn:6}, we obtain the nonlinear differential equations governing the tank dynamics:
\begin{equation}\label{Eq:Dyn:7}
    \begin{aligned}
    \frac{dh_1}{dt} &= -\frac{a_1}{A_1} \sqrt{2g h_1} + \frac{a_3}{A_1} \sqrt{2g h_3} + \gamma_1 \frac{k_1}{A_1} v_1, \\
    \frac{dh_2}{dt} &= -\frac{a_2}{A_2} \sqrt{2g h_2} + \frac{a_4}{A_2} \sqrt{2g h_4} + \gamma_2 \frac{k_2}{A_2} v_2, \\
    \frac{dh_3}{dt} &= -\frac{a_3}{A_3} \sqrt{2g h_3} + \frac{(1 - \gamma_2) k_2}{A_3} v_2, \\
    \frac{dh_4}{dt} &= -\frac{a_4}{A_4} \sqrt{2g h_4} + \frac{(1 - \gamma_1) k_1}{A_4} v_1.
    \end{aligned}
\end{equation}

Stacking the tank levels as a state vector \( \mathbf{h} = [h_1, h_2, h_3, h_4]^\top \) and control inputs as \( \mathbf{v} = [v_1, v_2]^\top \), the overall system can be expressed in compact nonlinear form as:
\begin{equation}\label{Eq:Dyn:8}
    \dot{\mathbf{h}} = f(\mathbf{h}, \mathbf{v}, \bar{\gamma}),
\end{equation}
where \( f \) is a structured nonlinear function incorporating the valve settings \( \bar{\gamma} = [\gamma_1, \gamma_2]^\top \), tank parameters \( A_i, a_i \), and pump gains \( k_1, k_2 \).

The model and control of the quadruple-tank process are analyzed at two operating points: $P_-$, where the system exhibits minimum-phase behavior, and $P_+$, where it demonstrates nonminimum-phase characteristics.

The nonlinear model in \eqref{Eq:Dyn:7} can be approximated by a linear system by introducing the deviation variables $x_i = h_i - h_i^0$ and $u_i = v_i - v_i^0$, where $h_i^0$ and $v_i^0$ denote the operating points of the water levels and input voltages, respectively. Using the standard state-space representation:
\begin{align}\label{Eq:Dyn:9}
    \dot{x} = Ax + Bu, \quad y = Hx,
\end{align}
the complete linearized model is expressed as:
\begin{align}\label{Eq:Dyn:10}
\begin{aligned}
    \frac{dx}{dt} &=
    \begin{bmatrix}
    -\dfrac{1}{\tau_1} & 0 & \dfrac{A_3}{A_1 \tau_3} & 0 \\
    0 & -\dfrac{1}{\tau_2} & 0 & \dfrac{A_4}{A_2 \tau_4} \\
    0 & 0 & -\dfrac{1}{\tau_3} & 0 \\
    0 & 0 & 0 & -\dfrac{1}{\tau_4}
    \end{bmatrix} x +
    \begin{bmatrix}
    \dfrac{\gamma_1 k_1}{A_1} & 0 \\
    0 & \dfrac{\gamma_2 k_2}{A_2} \\
    0 & \dfrac{(1 - \gamma_2) k_2}{A_2} \\
    \dfrac{(1 - \gamma_1) k_1}{A_1} & 0
    \end{bmatrix} u, \\
    y &= \begin{bmatrix}
    k_c & 0 & 0 & 0 \\
    0 & k_c & 0 & 0
    \end{bmatrix} x,
\end{aligned}
\end{align}
where $x \in \mathbb{R}^4$ is the state vector, $u \in \mathbb{R}^2$ is the input vector, and $y \in \mathbb{R}^2$ is the output vector. The matrices $(A, B, C)$ define the linearized system dynamics. Here, $k_c$ is the sensor gain, and the time constant $\tau_i$ of each tank depends on the nominal level $h_i^0$ and physical parameters $a_i$, $A_i$, and $g$, given by:
\begin{equation}\label{Eq:Dyn:11}
    \tau_i = \frac{A_i}{a_i} \sqrt{\frac{2 h_i^0}{g}}, \quad \forall i = 1, \dots, 4.
\end{equation}
The corresponding transfer function matrix is:
\begin{align}
    G(s) &= H(sI - A)^{-1}B \nonumber \\
    &=
    \begin{bmatrix}
    \dfrac{\gamma_1 c_1}{1 + s\tau_1} & \dfrac{(1 - \gamma_2)c_1}{(1 + s\tau_3)(1 + s\tau_1)} \\
    \dfrac{(1 - \gamma_1)c_2}{(1 + s\tau_1)(1 + s\tau_2)} & \dfrac{\gamma_2 c_2}{1 + s\tau_2}
    \end{bmatrix}, \quad
    c_p = \dfrac{\tau_p k_p k_c}{A_p}, \quad p \in \{1, 2\}.
    \label{Eq:Dyn:12}
\end{align}

\section{Analysis of Nonlinear Model}\label{Sec:3}
This section analyzes the characteristics of the nonlinear model \eqref{Eq:Dyn:7} of the quadruple-tank system. We focus on the derivation of nonlinear zero dynamics and the conditions for stationary operating points.

Given an initial state $(h_1^0, \dots, h_4^0)$ and an input pair $(v_1, v_2)$ such that the outputs remain identically zero, i.e., $h_1(t) = h_2(t) = 0$ for all $t \geq 0$, the internal dynamics that evolve under these constraints are known as the \textit{zero dynamics}. For the nonlinear model \eqref{Eq:Dyn:7}, these constrained dynamics reduce to:
\begin{equation}\label{Eq:NonDyn:13}
    \begin{aligned}
    \frac{dh_3}{dt} &= -\frac{a_3}{A_3} \sqrt{2g h_3} - \frac{(1 - \gamma_2) a_4}{\gamma_2 A_3} \sqrt{2g h_4}, \\
    \frac{dh_4}{dt} &= -\frac{(1 - \gamma_1) a_3}{\gamma_1 A_4} \sqrt{2g h_3} - \frac{a_4}{A_4} \sqrt{2g h_4}.
    \end{aligned}
\end{equation}
The characteristic equation of the linearized dynamics in \eqref{Eq:NonDyn:13} is:
\begin{equation*}
    (1 + s \tau_3)(1 + s \tau_4) - \frac{(1 - \gamma_1)(1 - \gamma_2)}{\gamma_1 \gamma_2} = 0,
\end{equation*}
where $\tau_3$ and $\tau_4$ are the time constants corresponding to tanks 3 and 4, as previously defined. The roots of this equation correspond to the transmission zeros of $G(s)$ (see Equation (7)). From this relation, it follows that:
\begin{itemize}
    \item the system is non-minimum phase when $0 < \gamma_1 + \gamma_2 < 1$,
    \item and minimum phase when $1 < \gamma_1 + \gamma_2 < 2$.
\end{itemize}
This outcome aligns with the intuition that linear approximation and zero dynamics computation commute. Some nonlinear control methods, such as feedback linearization, require that the zero dynamics be stable and thus cannot be applied globally in non-minimum-phase regimes. Since the zero dynamics in this system can become unstable depending on valve settings, the quadruple-tank serves as a useful benchmark to illustrate how sensitive nonlinear controllers are to the stability of zero dynamics.

To determine the steady-state levels $(h_1^0, h_2^0)$ corresponding to constant inputs $(v_1^0, v_2^0)$, we examine the system at equilibrium. From \eqref{Eq:Dyn:7}, the following steady-state equalities hold:
\begin{align*}
    \frac{a_3}{A_3} \sqrt{2g h_3^0} &= \frac{(1 - \gamma_2) k_2}{A_3} v_2^0, \\
    \frac{a_4}{A_4} \sqrt{2g h_4^0} &= \frac{(1 - \gamma_1) k_1}{A_4} v_1^0.
\end{align*}
Substituting into the expressions for $h_1$ and $h_2$ yields:
\begin{align}\label{Eq:NonDyn:14}
    \begin{aligned}
    \frac{a_1}{A_1} \sqrt{2g h_1^0} &= \frac{\gamma_1 k_1}{A_1} v_1^0 + \frac{(1 - \gamma_2) k_2}{A_3} v_2^0, \\
    \frac{a_2}{A_2} \sqrt{2g h_2^0} &= \frac{(1 - \gamma_1) k_1}{A_4} v_1^0 + \frac{\gamma_2 k_2}{A_2} v_2^0.
    \end{aligned}
\end{align}
It follows that a unique pair of constant control inputs $(v_1^0, v_2^0)$ exists to produce the steady-state water levels $(h_1^0, h_2^0)$ if and only if the following matrix is nonsingular:
\begin{equation*}
    \begin{bmatrix}
    \gamma_1 k_1 & (1 - \gamma_2) k_2 \\
    (1 - \gamma_1) k_1 & \gamma_2 k_2
    \end{bmatrix}
\end{equation*}
This matrix is singular if $\gamma_1 + \gamma_2 = 1$. The singularity reflects a physical degeneracy: in such cases, the flows into tanks 1 and 2 depend on the same linear combination of pump flows, making the outputs dependent. Consequently, the steady-state levels become interdependent as well.

\begin{remark}
    The valve ratios \( \gamma_1, \gamma_2 \) significantly influence the system's zero dynamics. When \( \gamma_1 + \gamma_2 < 1 \), the system exhibits minimum-phase behavior; otherwise, it becomes non-minimum-phase. This property underpins many of the challenges in control design discussed later. Notably, the condition $\gamma_1 + \gamma_2 = 1$ corresponds to the scenario where the transfer matrix $G(s)$ has a zero at the origin.
\end{remark} 

\section{Multivariable Zero}\label{Sec:4}
This section examines the zeros of the transfer matrix $G(s)$ in \eqref{Eq:Dyn:12}, emphasizing their physical significance based on the valve settings $\gamma_1$ and $\gamma_2$. The zeros are given by the roots of the numerator of $G(s)$:
\begin{equation}\label{Eq:Zero:15}
    \det G(s) = \frac{c_1 c_2}{\gamma_1 \gamma_2 \prod_{i=1}^4 (1 + s \tau_i)} \left[ (1 + s \tau_3)(1 + s \tau_4) - \frac{(1 - \gamma_1)(1 - \gamma_2)}{\gamma_1 \gamma_2} \right].
\end{equation}

Define the parameter
\begin{equation*}
    \eta := \frac{(1 - \gamma_1)(1 - \gamma_2)}{\gamma_1 \gamma_2}, \quad \eta \in (0, \infty).
\end{equation*}

The system behavior is classified as:
\begin{itemize}
    \item \textbf{Minimum-phase} if $1 < \gamma_1 + \gamma_2 < 2$,
    \item \textbf{Nonminimum-phase} if $0 < \gamma_1 + \gamma_2 < 1$,
    \item \textbf{Zero at the origin} if $\gamma_1 + \gamma_2 = 1$.
\end{itemize}

When $\eta$ is small, the zeros are close to $-1/\tau_3$ and $-1/\tau_4$. As $\eta \to \infty$, one zero approaches $+\infty$, the other $-\infty$. When $\eta = 1$, corresponding to $\gamma_1 + \gamma_2 = 1$, the system has a zero at the origin.

Examples:
\begin{itemize}
    \item $\gamma_1 + \gamma_2 = 1.30$ (e.g., $P_-$): minimum-phase,
    \item $\gamma_1 + \gamma_2 = 0.77$ (e.g., $P_+$): nonminimum-phase.
\end{itemize}

These dynamics have an intuitive interpretation. Let $q_i$ be the flow from pump $i$ for simplicity, and assume $q_1 = q_2$. The total flow to the upper tanks is $[2 - (\gamma_1 + \gamma_2)]q_1$ while the total flow to lower tanks is $(\gamma_1 + \gamma_2)q_1$. If $\gamma_1 + \gamma_2 > 1$, more flow reaches the lower tanks, simplifying control of $y_1$ and $y_2$ via $u_1$ and $u_2$. If $\gamma_1 + \gamma_2 = 1$, the flow is balanced between left and right tanks—this corresponds to a multivariable zero at the origin and leads to the most challenging control scenario.

In multivariable systems, unlike scalar ones, not only the location but also the direction of a zero matters. The direction of a zero $z$ is defined as a unit vector $\psi \in \mathbb{R}^2$ satisfying $\psi^\top G(z) = 0$. For $G(s)$ in \eqref{Eq:Dyn:12} and $z > 0$, this condition yields:
\begin{equation}\label{Eq:Zero:16}
    \begin{bmatrix}
        \psi_1 & \psi_2
    \end{bmatrix}
    \begin{bmatrix}
        \dfrac{\gamma_1 c_1}{1 + z \tau_1} & \dfrac{(1 - \gamma_2) c_1}{(1 + z \tau_3)(1 + z \tau_1)} \\
        \dfrac{(1 - \gamma_1) c_2}{(1 + z \tau_1)(1 + z \tau_2)} & \dfrac{\gamma_2 c_2}{1 + z \tau_2}
    \end{bmatrix}
    =
    \begin{bmatrix}
        0 & 0
    \end{bmatrix}.
\end{equation}
From this, it follows that $\psi_1, \psi_2 \neq 0$, implying the zero is associated with both outputs. Solving \eqref{Eq:Zero:16} for $\psi_1/\psi_2$ when $\gamma_2$ is known gives:
\begin{equation}\label{Eq:Zero:17}
    \frac{\psi_1}{\psi_2} = -\frac{1 - \gamma_1}{\gamma_1} \cdot \frac{c_2(1 + z \tau_1)}{c_1(1 + z \tau_4)(1 + z \tau_2)}.
\end{equation}
This shows that if $\gamma_1$ is small, the zero mainly affects $y_1$; if $\gamma_1$ is close to 1, the influence is more on $y_2$. Thus, the direction is determined by the relative sizes of $\gamma_1$ and $\gamma_2$.

The Relative Gain Array (RGA) $\Lambda$ is used to assess input-output pairing suitability for decentralized control. For a $2 \times 2$ system, RGA is defined as $\Lambda = G(0) \circ [G^{-1}(0)]^\top$, where $\circ$ is the Hadamard product. For the quadruple-tank system, the scalar RGA value $\lambda$ is:
\begin{equation}\label{Eq:Zero:18}
    \lambda = \frac{\gamma_1 \gamma_2}{\gamma_1 + \gamma_2 - 1}.
\end{equation}
while the complete RGA matrix is:
\begin{equation}\label{Eq:Zero:19}
    \text{RGA } \Lambda =
    \begin{bmatrix}
    \dfrac{\gamma_1 \gamma_2}{\gamma_1 + \gamma_2 - 1} & \dfrac{-(1 - \gamma_1)(1 - \gamma_2)}{\gamma_1 + \gamma_2 - 1} \\
    \dfrac{-(1 - \gamma_1)(1 - \gamma_2)}{\gamma_1 + \gamma_2 - 1} & \dfrac{\gamma_1 \gamma_2}{\gamma_1 + \gamma_2 - 1}
    \end{bmatrix}
    \rightarrow
    \begin{bmatrix}
    \lambda & 1 - \lambda \\
    1 - \lambda & \lambda
    \end{bmatrix}
\end{equation}
This symmetric structure reflects the interaction between the inputs and outputs.

This expression depends solely on valve ratios $\gamma_1, \gamma_2$ and not on other system parameters. The pairing is generally considered effective when $0.67 < \lambda < 1.50$. If $\gamma_1 + \gamma_2 < 1$, then $\lambda < 0$, which indicates a problematic pairing. In this case, swapping the outputs $y_1 \leftrightarrow y_2$ yields a new transfer matrix $\tilde{G}(s)$ with RGA:
\begin{equation}\label{Eq:Zero:20}
    \tilde{\lambda} = \frac{(1 - \gamma_1)(1 - \gamma_2)}{1 - \gamma_1 - \gamma_2}.
\end{equation}
Hence, if $\gamma_1 + \gamma_2 < 1$, then $\tilde{\lambda} > 0$, making the alternative pairing favorable. This matches physical intuition, where better control is achieved by directing more flow to the lower tanks. Thus, both the location and direction of zeros, together with RGA-based pairing, are tightly linked to valve settings. These structural properties highlight why decentralized control can fail under poor pairing or zero dynamics instability.

\section{Decentralized Control and Stability Condition}\label{Sec:5}
We consider a LTI MIMO plant modeled by the linear input-output relationship:
\begin{equation}\label{Eq:Control:21}
    y(s) = G(s)u(s), \quad \longrightarrow \quad \begin{bmatrix}
        y_1(s) \\
        \vdots \\
        y_p(s)
    \end{bmatrix} = \begin{bmatrix}
    G_{1,1} & \cdots & G_{1,m} \\
    \vdots & \ddots & \vdots \\
    G_{p,1} & \cdots & G_{p,m} 
    \end{bmatrix} \begin{bmatrix}
        u_1(s) \\
        \vdots \\
        u_m(s)
    \end{bmatrix}
\end{equation}
where $y(s) = [y_1(s),\dots,y_p(s)]^\top$ and $u(s) = [u_1(s),\dots,u_m(s)]^\top$ are the Laplace transforms of the output and input signals, respectively. The matrix $G(s) \in \mathbb{C}^{p \times m}$ denotes the transfer function of the system. In this work, we assume a square plant ($p = m$) and that $G(s)$ is stable rational transfer function.

Our goal is to design a decentralized controller that ensures overall system stability—even in the presence of uncertainties—while achieving desired performance levels. This controller should rely only on local measurements and actions.

The decentralized design proceeds in two major steps:
\begin{enumerate}
    \item \textbf{Input-output pairing:} Identify suitable pairings between inputs and outputs. A standard approach is to evaluate the \textit{Relative Gain Array} (RGA) of the plant at steady state:
    \begin{equation*}
        \Lambda = G(0) \circ [G(0)^{-1}]^\top,
    \end{equation*}
    where elements of \( \Lambda \) close to 1 (i.e., \( \Lambda_{ij} \approx 1 \)) indicates a favorable pairing between $u_j$ and $y_i$ (See Section~\ref{Sec:4}). Incorrect pairing can result in poor performance or instability, even if each individual loop is internally stabilized. To assess whether the chosen pairing permits decentralized stabilization, the Niederlinski Index (NI) is often employed:
    \begin{equation}\label{Eq:Control:22}
        \text{NI} = \frac{\det G(0)}{\prod_{i=1}^{n} \diag\{G(0)\}}.
    \end{equation}
    A negative index (\( \text{NI} < 0 \)) indicates that the selected configuration is not stabilizable using decentralized control, necessitating a reevaluation of the pairings.
    
    However, caution is warranted: both RGA and NI are based on steady-state information and may fail to reflect essential dynamics in systems with significant directional interactions (e.g., lower or upper triangular systems). In such cases, dynamic metrics like the PRGA offer better insight into pairing suitability by incorporating frequency-domain performance considerations. 
    
    \item \textbf{Diagonal controller design:} For the selected pairing, reorder \( G(s) \) via permutation matrices \( P_u \), \( P_y \) such that:
    \begin{equation*}
        \tilde{G}(s) = P_y G(s) P_u^\top
    \end{equation*}
    aligns the dominant pairings along the diagonal. Then, design each local loop \( C_i(s) \) independently to stabilize and regulate its assigned channel.
\end{enumerate}

This allows the decentralized controller to be expressed as a diagonal matrix:
\begin{equation*}
    C(s) = \diag\{C_1(s), C_2(s), \dots, C_m(s)\}
\end{equation*}
where each \( C_i(s) \) uses only local feedback \( y_i \mapsto u_i \), such that the closed-loop system $u(s) = -C(s) y(s)$ ensures internal stability and meets specified performance criteria (e.g., reference tracking, disturbance rejection, robustness).

It is important to note that the last two objectives may sometimes be contradictory, especially in the presence of strong loop interactions. The closed-loop sensitivity and complementary sensitivity functions are defined as:
\begin{align}\label{Eq:Control:23}
    \begin{aligned}
    S(s) &= \left[I + G(s)C(s)\right]^{-1}, \\
    T(s) &= G(s)C(s)\left[I + G(s)C(s)\right]^{-1}.
    \end{aligned}
\end{align}

\subsection{Robust Stability under Decentralized Control}

In practice, model uncertainty is inevitable—due to parameter variations, unmodeled dynamics, or simplifications in modeling. To analyze the robustness of decentralized control in the presence of such uncertainties, we decompose the plant transfer matrix \( G(s) \) into two parts:
\begin{equation}\label{Eq:Control:24}
    G(s) = G_D(s) + G_M(s),
\end{equation}
where \( G_D(s) \) is the nominal (diagonal) component, representing the decoupled dynamics assumed during controller design, and \( G_M(s) \) captures the coupling and uncertainty, modeled as a perturbation or mismatch between the actual plant and its decentralized approximation.

The stability of the closed-loop system can be analyzed using the \textit{small gain theorem}, which asserts that the feedback interconnection of two stable systems is internally stable if the product of their gains is strictly less than one.

\begin{lemma}[Veselý and Harsányi, 2008]
    Let \( G(s) \) be a stable LTI MIMO system, and let \( C(s) \) be a decentralized controller such that \( G_D(s)C(s) \) is internally stable. Then the overall closed-loop system is stable if:
    \begin{equation}\label{Eq:Control:25}
        \left\| G_D^{-1}(s) W(s) \right\| \cdot \left\| G_M(s) \right\| < 1,
    \end{equation}
    or equivalently,
    \begin{equation}\label{Eq:Control:26}
        \left\| G_D^{-1}(s) W(s) \right\| < \frac{1}{\left\| G_M(s) \right\|},
    \end{equation}
    where \( W(s) = C^{-1}(s) + G_D(s) \) is the open-loop return function of the nominal system.
\end{lemma}

This condition can be alternatively expressed in terms of the nominal closed-loop transfer function:
\begin{equation}\label{Eq:Control:27}
    \left\| G_D^{-1}(s) T_D(s) \right\| < M_0 := \frac{1}{\left\| G_M(s) \right\|},
\end{equation}
where $T_D(s) = G_D(s) C(s) \left[I + G_D(s) C(s)\right]^{-1}$ is the nominal closed-loop transfer matrix.

The condition in \eqref{Eq:Control:27} is valid for stable systems with no right-half-plane (RHP) transmission zeros, including both minimum and non-minimum phase configurations. However, it may become overly conservative in low-frequency regimes, particularly when \( \|T_D(s)\| \approx 1 \), which limits its utility in practice.

To alleviate this conservatism, a less restrictive condition is given by the following result:

\begin{lemma}[Skogestad and Postlethwaite, 2009]
    Let \( G(s) \) be a stable square MIMO system with a decentralized controller \( C(s) \), and assume that neither \( G(s) \) nor its nominal approximation \( G_D(s) \) has RHP transmission zeros. Define the nominal sensitivity function:
    \[
    S_D(s) = \left(I + G_D(s)C(s)\right)^{-1}
    \]
    and the normalized interaction matrix:
    \[
    E(s) = G(s) - G_D(s) G^{-1}(s) = G_M(s) G^{-1}(s).
    \]
    Then the closed-loop system is stable if the inverse of \( I - E(s) S_D(s) \) exists. A sufficient condition for robust stability is:
    \begin{equation}\label{Eq:Control:28}
        \left\| E(s) S_D(s) \right\| < 1,
        \quad \text{or equivalently} \quad
        \left\| G^{-1}(s) S_D(s) \right\| < \frac{1}{\left\| G_M(s) \right\|}.
    \end{equation}
\end{lemma}

Both conditions—\eqref{Eq:Control:27} and \eqref{Eq:Control:28}—can be employed to certify the internal stability of the closed-loop system under decentralized control in the presence of model uncertainty. The choice of which condition to apply depends on available system information and the conservativeness acceptable in a given application.

\begin{remark}
    Stability of individual loops \( G_{ii}(s)C_i(s) \) does not imply overall closed-loop stability when the off-diagonal terms of \( G(s) \) are non-negligible. That is, the condition
    \[
        \|G(s) - \diag\{G_{11}(s), \dots, G_{mm}(s)\}\| \ll 1
    \]
    is often necessary for decentralized design to be valid. Hence, evaluating multivariable loop interactions is essential even in decentralized architectures.
\end{remark}
\begin{figure}
    \centering
    \includegraphics[width=0.65\linewidth]{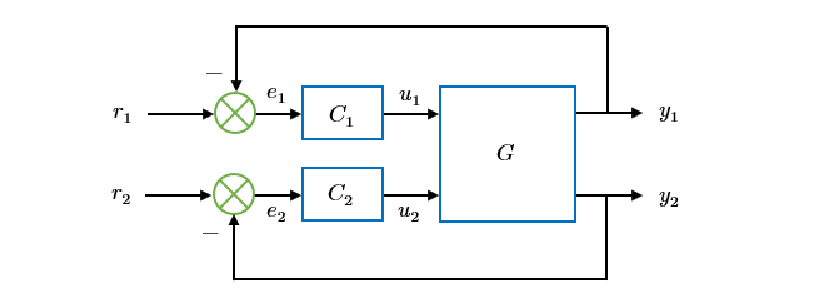}
    \caption{Structure of decentralized control architecture highlighting interactions between local controllers $C_1$ and $C_2$.}
    \label{Fig2}
\end{figure}

\subsection{Decentralized PI Control and Performance Limits}
To regulate the water levels in the lower tanks, we implement a decentralized PI controller structure. The objective is to design local control loops such that each input \( u_i \) regulates the corresponding output \( y_i \) based solely on local error signals \( e_i = r_i - y_i \). The overall decentralized control law is defined as:
\begin{equation}\label{Eq:Control:29}
    u = \diag\{C_1, C_2\} e, \quad e = [e_1, e_2]^\top,
\end{equation}
where each \( C_p(s) \) is a PI controller of the form:
\begin{equation}\label{Eq:Control:30}
    C_p(s) = K_p\left(1 + \frac{1}{\tau_{i,p} s} \right), \quad p = 1, 2.
\end{equation}
The gains are manually tuned based on simulations of the linearized physical model. For the minimum-phase setting, controller tuning is straightforward. Using the parameters:
\[
    (K_1, \tau_{i,1}) = (3.0, 30), \quad (K_2, \tau_{i,2}) = (2.7, 40),
\]
the system achieves good closed-loop performance. In contrast, tuning for the nonminimum-phase configuration is considerably more difficult. Even with stabilizing settings:
\[
(K_1, \tau_{i,1}) = (1.5, 110), \quad (K_2, \tau_{i,2}) = (-0.12, 220),
\]
the closed-loop response is much slower. The settling time is nearly 10 times longer than in the minimum-phase case. This is partly due to the reduced gain \( K_2 \), and fundamentally to the presence of a right-half-plane zero.

Control performance is highly sensitive to the choice of output-input pairing. As discussed in Section~\ref{Sec:4}, the original pairing yields an RGA value \( \lambda = -0.64 \), whereas permuting \( y_1 \leftrightarrow y_2 \) improves the RGA to \( \tilde{\lambda} = 1.64 \). Although this permutation does not change the location of the RHP zero, it significantly improves performance due to better loop decoupling. Centralized multivariable control (e.g., \( H_\infty \) methods) has also been tested on the quadruple-tank system. In the minimum-phase case, such methods provide limited improvement over decentralized PI control. For the nonminimum-phase case, however, \( H_\infty \) controllers achieve 30--40\% faster settling times compared to decentralized control. Interestingly, the optimal centralized controller adopts an anti-diagonal structure—consistent with the RGA-motivated permutation.

Fundamental performance limits are dictated by the location and direction of system zeros. According to~\cite{R1}, if the RHP zero lies close to the origin, and there is fixed undershoot in \( y_1 \) and strong interaction to \( y_2 \), then any linear controller will yield a large settling time. This limit is unavoidable and holds regardless of controller structure. In contrast, the minimum-phase setting theoretically allows arbitrarily tight regulation using purely decentralized controllers.

\section{Distributed Estimation}
In this section, we design and analyze a distributed observer for reconstructing the full state of a system from local measurements and limited communication. We consider the following continuous-time linear time-invariant (LTI) system:
\begin{equation}\label{Eq:Estimate:31}
    \dot{x} = Ax + Bu, \quad y = Hx
\end{equation}
where \(x \in \mathbb{R}^n\) is the system state, \(u \in \mathbb{R}^m\) is the control input, and \(y \in \mathbb{R}^p\) is the measured output. The output matrix \(H \in \mathbb{R}^{p \times n}\) is decomposed as \(H = \mathrm{col}(H_1, H_2, \dots, H_N)\), where each local matrix \(H_i \in \mathbb{R}^{p_i \times n}\) satisfies \(\sum_{i=1}^N p_i = p\). The individual measurement \(y_i = H_i x \in \mathbb{R}^{p_i}\) is available only to node \(i\), forming a localized sensing structure.

The distributed estimation problem seeks to construct a network of local observers, each utilizing only its local measurement and communicating with neighboring nodes over a given graph. This graph is assumed to be a strongly connected, directed network \(\mathcal{G} = (\mathcal{N}, \mathcal{E}, \mathcal{A})\), with node set \(\mathcal{N} = \{1,\dots,N\}\), edge set \(\mathcal{E} \subset \mathcal{N} \times \mathcal{N}\), and weighted adjacency matrix \(\mathcal{A} = [a_{ij}] \in \mathbb{R}^{N \times N}\), where \(a_{ij} > 0\) if node \(j\) communicates with node \(i\). The Laplacian matrix is defined as \(\mathcal{L} = D - \mathcal{A}\), where \(D = \mathrm{diag}(d_i)\), with \(d_i = \sum_{j=1}^N a_{ij}\).
\begin{figure}[h!]
    \centering
    \includegraphics[width=0.5\linewidth]{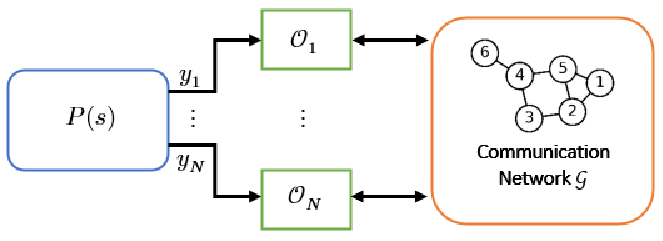}
    \caption{Scheme of distributed estimation over $N$ output $y_N$ and observer $O_N$}
    \label{Fig3}
\end{figure}
\begin{definition}
A central goal is to design local observers \(\hat{x}_i\) such that, despite relying on partial measurements, all state estimates asymptotically converge to the true state. This property is known as \textit{omniscience}. Formally, the distributed observer achieves omniscience if
\begin{equation}\label{Eq:Estimate:32}
    \lim_{t \to \infty} \|\hat{x}_i(t) - x(t)\| = 0, \quad \forall i \in \mathcal{N}.
\end{equation}
\end{definition}

The objective is to design a distributed observer where each node maintains a local observer of the form:
\begin{equation}\label{Eq:Estimate:33}
    \dot{\hat{x}}_i = A\hat{x}_i + Bu + L_i (y_i - H_i \hat{x}_i) + \alpha M_i^{-1}(\kappa_i) \sum_{j \in \mathcal{N}_i} a_{ij} (\hat{x}_j - \hat{x}_i), \quad i \in \mathcal{N}
\end{equation}
where \(\alpha > 0\) is a global coupling gain, \(L_i \in \mathbb{R}^{n \times p_i}\) is the local injection gain, and \(M_i \in \mathbb{R}^{n \times n}\) is a weighting matrix. The set \(\mathcal{N}_i\) denotes the neighbors of node \(i\) in the graph.

To design \(L_i\) and \(M_i\), we employ an observability decomposition. Let \(T_i \in \mathbb{R}^{n \times n}\) be an orthonormal matrix such that
\[
    T_i^\top A T_i = \begin{bmatrix}
        A_{\text{io}} & 0 \\
        A_{\text{ir}} & A_{\text{iu}}
    \end{bmatrix}, \qquad H_i T_i = \begin{bmatrix}
        H_{\text{io}} & 0
    \end{bmatrix}
\]
where the pair \((H_{\text{io}}, A_{\text{io}})\) is observable, and \(A_{\text{iu}} \in \mathbb{R}^{(n - \nu_i) \times (n - \nu_i)}\) represents the unobservable modes, with \(\nu_i\) being the observability index of \((H_i, A)\). The matrix \(T_i\) can be chosen such that the unobservable subspace is captured by the second block of columns.

Using this decomposition, we design the gains $L_i$ and $M_i(\kappa_i)$ as:
\begin{align}\label{Eq:Estimate:34}
    L_i = T_i \begin{bmatrix}
        L_{\text{io}} \\
        0
    \end{bmatrix}, \quad
    M_i(\kappa_i) = T_i \begin{bmatrix}
        \kappa_i M_{\text{io}} & 0 \\
        0 & I_{\nu_i}
    \end{bmatrix} T_i^\top
\end{align}
where \(\kappa_i \geq 1\) is a design parameter. The matrix \(L_{\text{io}}\) is chosen such that \(A_{\text{io}} - L_{\text{io}}H_{\text{io}}\) is Hurwitz and \(M_{\text{io}}\succ 0\) is the solution to the Lyapunov equation:
\begin{equation}\label{Eq:Estimate:35}
    (A_{\text{io}} - L_{\text{io}} H_{\text{io}})^\top M_{\text{io}} + M_{\text{io}}(A_{\text{io}} - L_{\text{io}} H_{\text{io}}) = -I_{\nu_i}
\end{equation}
ensuring that the local error subsystem is stable. This design guarantees that the estimation error at each node converges exponentially to zero for sufficiently large \(\alpha\) and appropriate choice of \(\kappa_i\).

To analyze the structure of the transformed error dynamics, consider any vector \( w \in \mathbb{R}^n \), and for each node \( i \in \mathcal{N} \), define the coordinate partition induced by the orthonormal matrix \( T_i \) as:
\begin{equation}\label{Eq:Estimate:36}
    T_i^\top w = 
    \begin{bmatrix}
        u_i \\
        v_i
    \end{bmatrix}, \quad u_i \in \mathbb{R}^{n - \nu_i},\ v_i \in \mathbb{R}^{\nu_i}.
\end{equation}
Since \( T_i \) is orthonormal, it preserves the norm, so \( \|w\|^2 = \|u_i\|^2 + \|v_i\|^2 \).

Now consider the weighted transformed system matrix \( M_i(\kappa_i)(A - L_i H_i) \). Using the structure of \( L_i \), \( M_i \), and the observability decomposition, we compute:
\begin{align*}
    M_i(\kappa_i)(A - L_i H_i)
    &= T_i
    \begin{bmatrix}
        \kappa_i M_{\text{io}} & 0 \\
        0 & I_{\nu_i}
    \end{bmatrix}
    T_i^\top (A - L_i H_i) T_i T_i^\top \\
    &= T_i
    \begin{bmatrix}
        \kappa_i M_{\text{io}} & 0 \\
        0 & I_{\nu_i}
    \end{bmatrix}
    \left(
    T_i^\top A T_i - T_i^\top L_i H_i T_i
    \right)
    T_i^\top \\
    &= T_i
    \begin{bmatrix}
        \kappa_i M_{\text{io}} & 0 \\
        0 & I_{\nu_i}
    \end{bmatrix}
    \begin{bmatrix}
        A_{\text{io}} - L_{\text{io}} H_{\text{io}} & 0 \\
        A_{\text{ir}} & A_{\text{iu}}
    \end{bmatrix}
    T_i^\top.
\end{align*}

This shows how the observer design leverages the block structure from the observability decomposition, allowing separate treatment of the detectable and undetectable modes.

Letting the estimation error be \(e_i = \hat{x}_i - x\), the local error dynamics are:
\begin{equation}\label{Eq:Estimate:37}
    \dot{e}_i = (A - L_i H_i) e_i + \alpha M_i^{-1} \sum_{j \in \mathcal{N}_i} a_{ij}(e_j - e_i)
\end{equation}
The global error system becomes:
\begin{equation}\label{Eq:Estimate:38}
    \dot{e} = \left[ \bar{\Lambda} - \alpha \bar{M}(\mathcal{L} \otimes I_n) \right]e
\end{equation}
with \(\bar{\Lambda} = \mathrm{diag}(A - L_1 H_1, \dots, A - L_N H_N)\) and \(\bar{M} = \mathrm{diag}(M_1, \dots, M_N)\). The system achieves omniscience, i.e., \(\lim_{t \to \infty} \hat{x}_i(t) = x(t)\), if and only if the global error system is asymptotically stable.

\begin{theorem}
    Suppose that the pair \( (A, H) \) and the communication graph \( \mathcal{G} \) satisfy the required detectability and connectivity conditions. Then, for each node \( i \in \mathcal{N} \), the distributed observer achieves omniscience asymptotically:
    \begin{equation*}
        \lim_{t \to \infty} \| \hat{x}_i(t) - x(t) \| = 0.
    \end{equation*}
    This holds provided that the gains \( \kappa_i \geq 1 \) and the coupling gain \( \alpha > 0 \) are selected such that:
    \begin{align}\label{Eq:Estimate:39}\begin{aligned}
        \left( \kappa_i - \frac{\beta}{\theta(\bar{\epsilon})} \right)
        \left( \alpha - \frac{\bar{\beta}}{2 \lambda_2} \right)
        &> \frac{\bar{\beta}^2 N \bar{\epsilon}^2}{2 \lambda_2 \theta(\bar{\epsilon})}, \quad \forall i \in \mathcal{N} \\
        \alpha &> \frac{\bar{\beta}}{2 \lambda_2},
        \end{aligned}
    \end{align}
    where
    \[
        \theta(\bar{\epsilon}) = \frac{1}{2} \left( 1 - \left(1 - \frac{\bar{\epsilon}^2}{2} \right)^2 \right),
    \]
    and constants \(\beta_i := 2\|A_{\emph{ir}}\|^2 + \|A_{\emph{iu}}^\top + A_{\emph{iu}}\|\), \(\beta = \sum_{i=1}^N \beta_i\), and \(\bar{\beta} := \max_{i \in \mathcal{N}} \beta_i\).
\end{theorem}

\begin{remark}
    The parameters \( \bar{\epsilon}, N, \lambda_2, \bar{\beta}, \beta \) are global quantities that require knowledge of the graph Laplacian \(\mathcal{L}\) and the full output matrix \( H \). While the structure is distributed, the selection of \(\kappa_i\) and \(\alpha\) requires centralized information. If each node has bounds on \(H_i\) and \(N\), local design is possible. Otherwise, adaptive gain selection is an alternative (see Section~IV).
\end{remark}
\begin{remark}
    The feasibility condition \eqref{Eq:Estimate:39} defines a region in the \((\kappa_i, \alpha)\)-plane where omniscience is guaranteed. For fixed \(\alpha > \frac{\bar{\beta}}{2\lambda_2}\), the parameter \(\kappa_i\) must satisfy both:
    \[
    \kappa_i > \frac{\beta}{\theta(\bar{\epsilon})}, \quad \kappa_i > \frac{\bar{\beta}^2 N \bar{\epsilon}^2}{2\lambda_2 \theta(\bar{\epsilon})(\alpha - \frac{\bar{\beta}}{2\lambda_2})}.
    \]
    The second expression defines a reciprocal curve. Thus, the feasible region lies above both the horizontal line and the curve. For any \(\alpha\) sufficiently large, a corresponding \(\kappa_i\) satisfying the condition always exists.
\end{remark}

\section{Numerical Results}
This section presents the simulation results that validate the proposed decentralized control and distributed estimation approaches for the quadruple-tank process. Three distinct numerical experiments are conducted: (i) the control of the physical quadruple-tank system under different phase conditions, (ii) the distributed observer simulation based on a low-dimensional Luenberger structure, and (iii) the distributed estimation over a larger interconnected system. The last two are intended to offer conceptual insight into distributed estimation strategies.
\begin{table}[h!]
    \centering
    \caption{Parameter of the laboratory-scale quadruple-tank}
    \label{table:params}
    \begin{tabular}{|c|c|c|}
    \toprule
    \textbf{Variable} & \textbf{Unit} & \textbf{Values} \\
    \midrule
    $A_1$, $A_3$ & cm\textsuperscript{2} & 28 \\
    $A_2$, $A_4$ & cm\textsuperscript{2} & 32 \\
    $a_1$, $a_3$ & cm\textsuperscript{2} & 0.071 \\
    $a_2$, $a_4$ & cm\textsuperscript{2} & 0.057 \\
    $k_c$ & V/cm & 0.5 \\
    $g$ & cm/s\textsuperscript{2} & 981 \\
    \bottomrule
    \end{tabular}
\end{table}
\begin{table}[h!]
    \centering
    \caption{Operating points of the minimum $P_-$ and non-minimum phase $P_+$ of the quadruple-tank process}
    \label{table:op_points}
    \begin{tabular}{|c|c|c|c|}
    \toprule
    \textbf{Variable} & \textbf{Unit} & $P_-$ & $P_+$ \\
    \midrule
    $(h_1^0, h_2^0)$ & cm & (12.4), (12.7) & (12.6), (13.0) \\
    $(h_3^0, h_4^0)$ & cm & (1.8), (1.4) & (4.8), (4.9) \\
    $(v_1^0, v_2^0)$ & V & (3.00), (3.00) & (3.15), (3.15) \\
    $(k_1, k_2)$ & cm\textsuperscript{3}/Vs & (3.33), (3.35) & (3.14), (3.29) \\
    $(\gamma_1, \gamma_2)$ & -- & (0.70), (0.60) & (0.43), (0.34) \\
    \bottomrule
    \end{tabular}
\end{table}
\begin{table}[h!]
    \centering
    \caption{Time-constant for the operating points $P_-$ and $P_+$}
    \label{table:time_constants}
    \begin{tabular}{|c|c|c|}
    \toprule
    \textbf{Variable} & $P_-$ & $P_+$ \\
    \midrule
    $(T_1, T_2)$ & (62.90) & (63.91) \\
    $(T_3, T_4)$ & (23.30) & (39.56) \\
    \bottomrule
    \end{tabular}
\end{table}

The linearized dynamics of the quadruple-tank process, described by \eqref{Eq:Dyn:10}, are simulated using the parameter sets in Tables~\ref{table:params}, \ref{table:op_points}, and \ref{table:time_constants}, given by:
\begin{equation}
    G_-(s) = 
    \begin{bmatrix}
    \dfrac{2.6}{1 + 62s} & \dfrac{1.5}{(1 + 23s)(1 + 62s)} \\
    \dfrac{1.4}{(1 + 30s)(1 + 90s)} & \dfrac{2.8}{1 + 90s}
    \end{bmatrix}, \quad
    G_+(s) = 
    \begin{bmatrix}
    \dfrac{1.5}{1 + 63s} & \dfrac{2.5}{(1 + 39s)(1 + 63s)} \\
    \dfrac{2.5}{1 + 91s} & \dfrac{1.6}{1 + 91s}
    \end{bmatrix}
\end{equation}
Two scenarios are considered: the minimum-phase configuration ($P_-$) and the non-minimum-phase configuration ($P_+$). The decentralized PI controllers are implemented with the following parameters: for $P_-$, $(K_1, T_1) = (3, 30)$ and $(K_2, T_2) = (2.7, 40)$; for $P_+$, $(K_1, T_1) = (1.5, 110)$ and $(K_2, T_2) = (-0.1, 220)$. The longer settling time in the non-minimum case reflects the inherent difficulty in controlling this unstable configuration.

The measurement outputs for each agent are modeled as:
\begin{equation}
    y_i = H_i x, \quad
    H_1 = \begin{bmatrix} k_c & 0 & 0 & 0 \end{bmatrix}, \quad
    H_2 = \begin{bmatrix} 0 & k_c & 0 & 0 \end{bmatrix}
\end{equation}
with $N=2$ agents communicating bidirectionally. The initial condition for the state vector is $x_0 = [8\;\;5\;\;-2\;\;1]^\top$, and the controller gain $\alpha = 6$ is used along with pump gains $k_1 = 3$ and $k_2 = 4.5$.

The simulation for the minimum-phase case demonstrates effective regulation of the tank levels, despite disturbances introduced through step changes in the setpoints at time instances $t = 100,\;200,\;300,\;350$ seconds. The system exhibits prompt convergence, as depicted in Fig.~\ref{F4b}, and the corresponding tracking error shown in Fig.~\ref{F4a} confirms transient performance within acceptable limits. In contrast, for the non-minimum-phase configuration, the system exhibits a slower response with significantly longer settling times and larger initial transients, as expected. This is evident in Fig.~\ref{F4d} and the tracking error in Fig.~\ref{F4d}. The challenges arise due to the internal unstable zero dynamics introduced by the flow ratios $\gamma_1 + \gamma_2 < 1$.

The effectiveness of the distributed estimation scheme is illustrated in Fig.\ref{F4c} for the minimum-phase case and Fig.\ref{F4f} for the non-minimum-phase configuration, utilizing a pair of distributed Luenberger observers with gains defined as:
\begin{equation*}
    T_1 = T_2 = 
    \begin{bmatrix}
    I_{r_2} & 0 \\
    0 & I_{r_1}
    \end{bmatrix}, \quad
    L_{1d} = \begin{bmatrix} 3 \\ 1 \end{bmatrix}, \quad
    L_{2d} = \begin{bmatrix} -1 \\ 3 \end{bmatrix}
\end{equation*}
\begin{equation*}
    M_{1d} = \begin{bmatrix} 0.5 & -0.5 \\ 0.5 & 1 \end{bmatrix}, \quad
    M_{2d} = \begin{bmatrix} 0.286 & -0.25 \\ -0.25 & 0.387 \end{bmatrix}.
\end{equation*}

In both scenarios, the state estimates produced by the distributed observers closely track the true system trajectories, despite each observer relying only on partial measurements and neighbor communication. The convergence of the estimated states validates the observer design and confirms that omniscience is achieved, even in the presence of nonlinear dynamics and limited sensing. This reinforces the viability of distributed estimation as a scalable solution for monitoring complex interconnected systems like the quadruple-tank process.

In addition to control simulations on the quadruple-tank system, we conduct two independent distributed observer simulations to conceptually validate the estimation framework. These simulations are not tied to the quadruple-tank dynamics but instead illustrate the behavior of distributed estimation under controlled experimental settings, as proposed in prior works.

The first experiment, based on the architecture in \cite{R12}, employs distributed Luenberger observers in a small-scale network. Each agent accesses a local output and communicates with neighbors to estimate the full system state. We compare two gain strategies: constant and adaptive coupling gain $\kappa_i$. As shown in Fig.\ref{F5a} and Fig.\ref{F5b}, both configurations achieve convergence to the true trajectory (indicated by black dashed lines), with the adaptive gain offering improved transient performance. This confirms the robustness of consensus-based estimation in basic multi-agent setups.

\begin{figure}[t!]
    \centering
    \subfloat[\label{F4a} ]{\includegraphics[width=.3\linewidth]{Graph/DEF1.eps}}
    \subfloat[\label{F4b} ]{\includegraphics[width=.3\linewidth]{Graph/DEF3.eps}}
    \subfloat[\label{F4c} ]{\includegraphics[width=.3\linewidth]{Graph/DEF5.eps}}\\
    \subfloat[\label{F4d} ]{\includegraphics[width=.3\linewidth]{Graph/DEF2.eps}}
    \subfloat[\label{F4e} ]{\includegraphics[width=.3\linewidth]{Graph/DEF4.eps}}
    \subfloat[\label{F4f} ]{\includegraphics[width=.3\linewidth]{Graph/DEF6.eps}}
    \caption{The error of the two parameters from the minimum-phase $P_-$ as (a) and the non-minimum $P_+$ as (b) using decentralized PI control; The two responses of the true output ($y$) from $P_-$ as (c) with 500s and $P_+$ as (d) with ten times longer settling time by 5000s with the same gains of control as designed; The response states of the distributed estimation of the $P_-$ (e) and $P_+$ (f) with the same initial conditions.}
    \label{F4}
\end{figure}

The second experiment, inspired by \cite{R13}, involves a higher-dimensional system with 
$n=6$ states and $N=4$ agents. Each agent observes a partial output vector and communicates over a directed graph. Despite the increased system complexity, the distributed observers collectively reconstruct the true state vector, as shown in Fig.~\ref{F6}. These results demonstrate the scalability and flexibility of the distributed estimation architecture, supporting its applicability to larger networked systems beyond the specific case of the quadruple-tank.

\begin{figure}[t!]
    \centering
    \subfloat[\label{F5a} With constant gain $\kappa_i$]{\includegraphics[width=.35\linewidth]{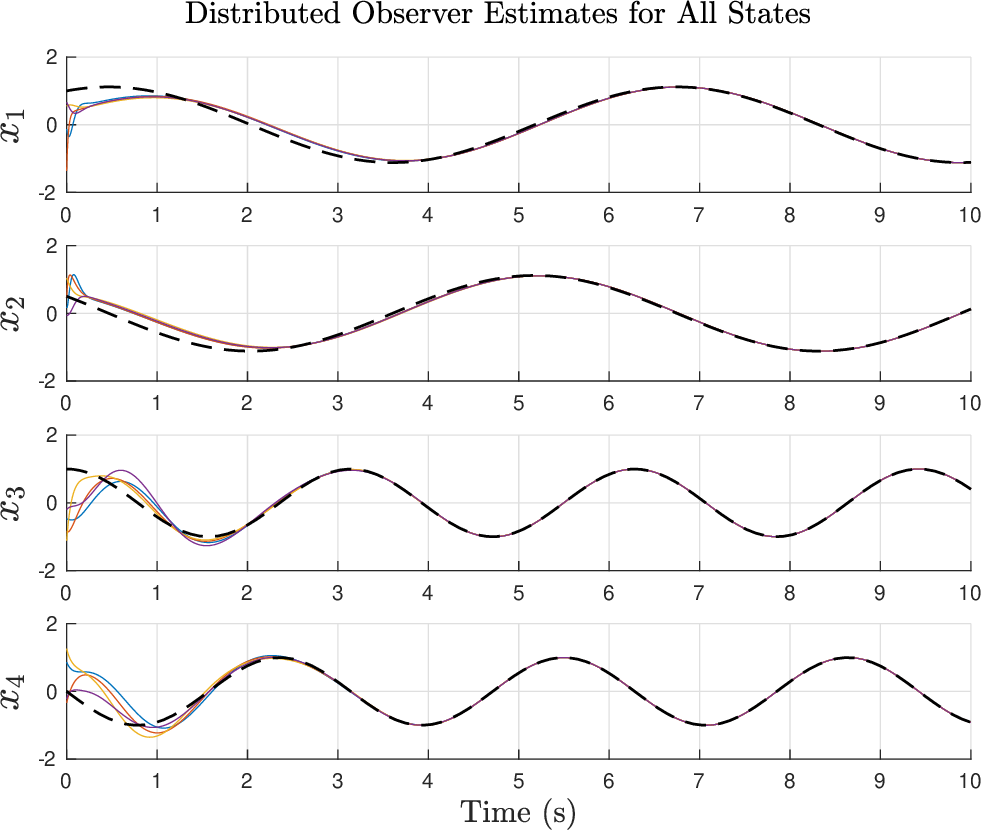}}\qquad
    \subfloat[\label{F5b} With adaptive gain $\kappa_i$]{\includegraphics[width=.35\linewidth]{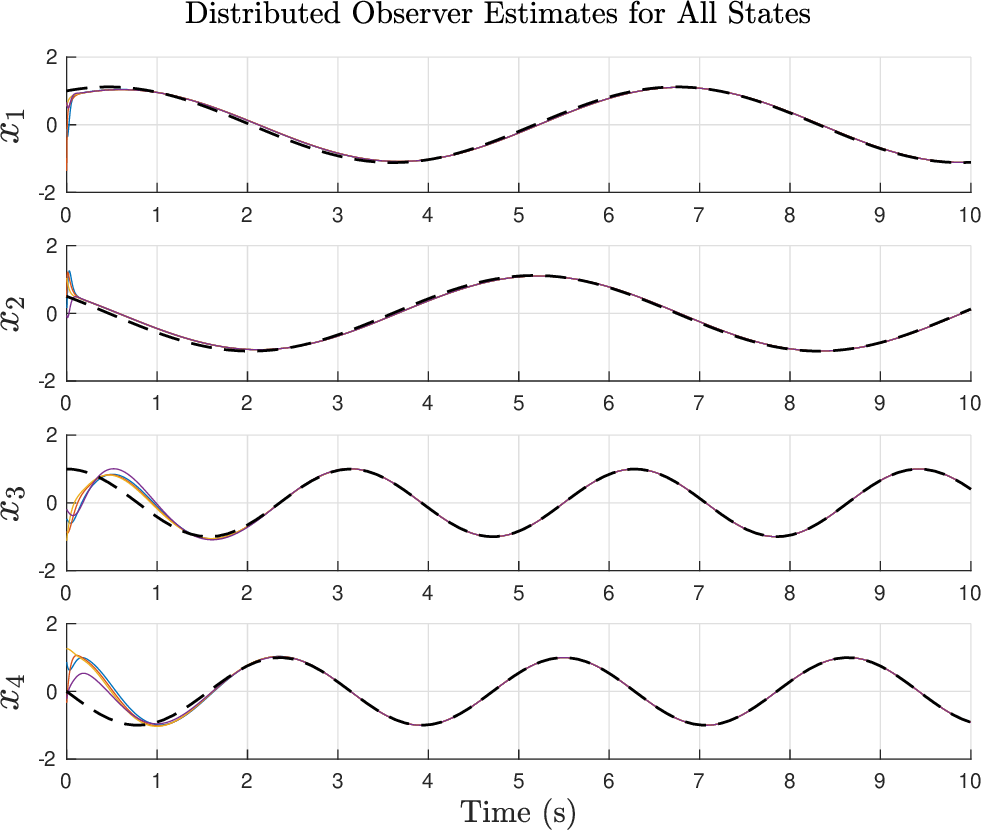}}
    \caption{Simulation results: trajectories of local estimates using \cite{R12}}
    \label{F5}
\end{figure}
\begin{figure}
    \centering
    \includegraphics[width=0.5\linewidth]{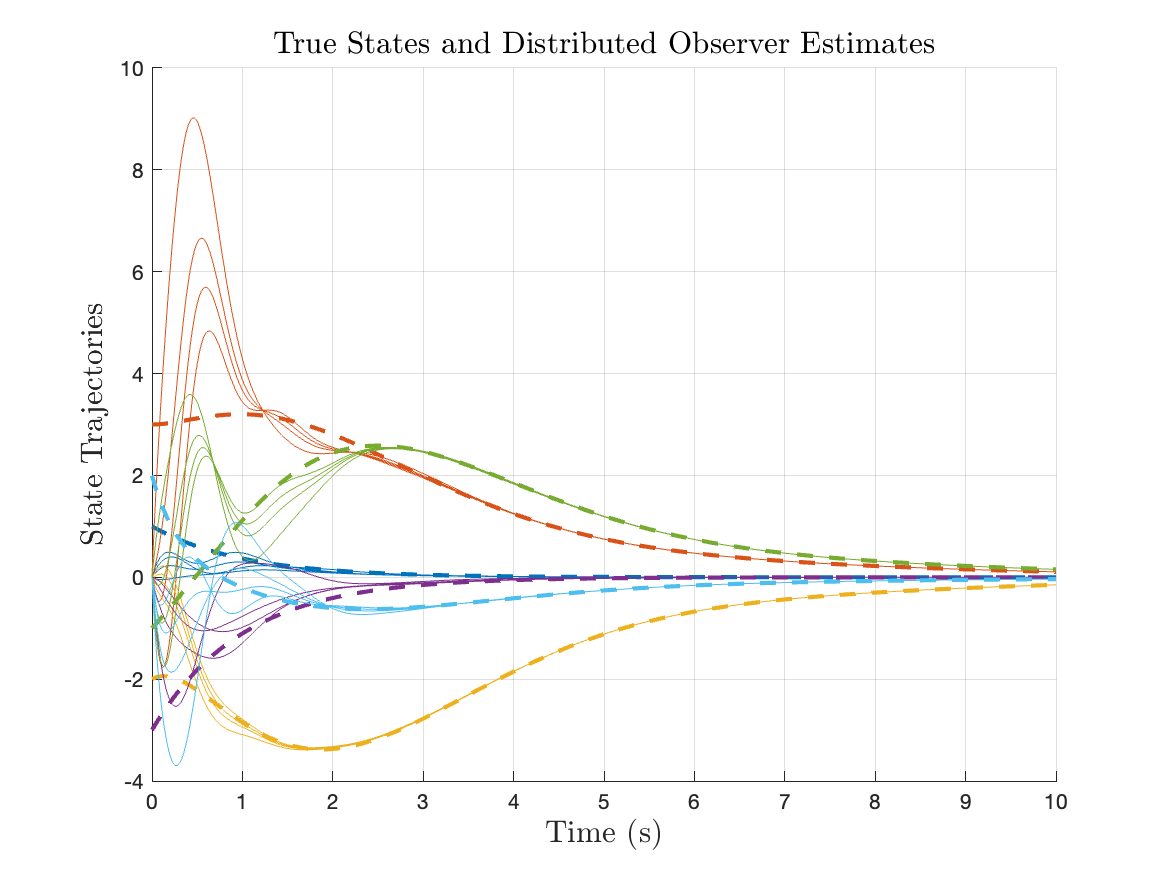}
    \caption{Simulation results: trajectories of local estimates using \cite{R13}}
    \label{F6}
\end{figure}

Overall, the simulation results demonstrate that:
1. The decentralized PI control successfully regulates the quadruple-tank system under both minimum and non-minimum configurations.
2. Distributed Luenberger observers, even with limited local measurements, can reconstruct the full state through communication and estimation logic.
3. The distributed estimation framework remains scalable, preserving convergence properties in larger, more interconnected systems.

These numerical insights support the viability of applying distributed estimation in practical networked control systems and highlight its compatibility with decentralized control strategies for nonlinear MIMO processes like the quadruple-tank system.

\section{Conclusion}
This work presented a systematic modeling, analysis, and control framework for the quadruple-tank process, with a focus on both decentralized control and distributed state estimation. The system exhibits fundamentally different behaviors under minimum-phase and non-minimum-phase configurations, which are determined by the flow-splitting parameters \(\gamma_1\) and \(\gamma_2\). Specifically, the condition \(\gamma_1 + \gamma_2 < 1\) results in a non-minimum-phase system, which is intrinsically harder to control due to the presence of right-half-plane (RHP) zeros. This was demonstrated through numerical scenarios, where decentralized PI control yielded acceptable performance in both configurations but with significantly longer settling times in the non-minimum-phase case—nearly ten times longer than the minimum-phase counterpart.

A decentralized PI control structure was implemented for each configuration. Although the controllers stabilized the system, the non-minimum-phase setting exhibited slower dynamics and more pronounced transient responses. This aligns with theoretical expectations that such systems are limited in performance due to fundamental constraints imposed by unstable zeros.

In parallel, a distributed state estimation architecture was developed based on local Luenberger observers and inter-node communication over a strongly connected graph. Each observer was designed using an observability decomposition and weighted error feedback from neighboring nodes. The proposed design guarantees omniscience under mild detectability and connectivity assumptions, and its effectiveness was verified through simulation. The observers successfully reconstructed the system state across both process configurations, including the more challenging non-minimum-phase case. The simulations confirmed that the estimation architecture is robust to localized sensing and capable of handling coupling dynamics induced by interconnection. While small transients appeared in the early phases due to initialization mismatch and communication lag, the estimators converged rapidly to the true system states.

Looking ahead, several directions merit further exploration. Enhancing the estimator architecture with distributed fault detection and fault-tolerant control capabilities could improve resilience. Adaptive and event-triggered communication strategies may also reduce bandwidth requirements while maintaining estimation quality. Finally, the proposed methods can be extended to more complex networked control systems with high-dimensional, heterogeneous dynamics and time-varying interaction graphs.

\section{Outlook and Future Directions}

The results presented in this work offer several avenues for future exploration, particularly at the intersection of distributed control, estimation, and learning.

One compelling direction is the development of \textit{distributed adaptive control} architectures. Unlike fixed-gain designs, adaptive controllers dynamically adjust parameters in real-time to account for system uncertainties or time-varying behaviors. Recent advances in this area include synchronization of agents in complex networks via adaptive feedback mechanisms \cite{FW1}, handling of heterogeneous actuator dynamics \cite{FW2}, and managing nonlinearities with full-state and input constraints in pure-feedback systems \cite{FW3}. For systems with delays or strong coupling, model reference adaptive control has also been explored to guarantee stability and convergence under structured communication graphs \cite{FW4}. Furthermore, high-order tuner strategies can be embedded to tackle structured disturbances across large interconnected systems \cite{FW5}. Applying such techniques to the quadruple-tank system would enable each node to autonomously tune its observer or controller using local performance metrics and limited neighbor communication.

A second direction is the integration of \textit{distributed fault-tolerant control} (FTC) and \textit{robust estimation}. In practical implementations, sensor degradation and actuator faults can severely degrade performance. To address this, active fault diagnosis methods can be layered onto decentralized control frameworks \cite{FW6}, enabling each node to isolate faults using only partial measurements. Several distributed FTC strategies have also been developed for consensus-based networks under actuator faults \cite{FW7}, sensor failures \cite{FW8}, and even simultaneous identification and reconfiguration \cite{FW9}. These approaches are often augmented with robust estimation tools to guarantee fault insensitivity or bounded performance in the presence of model uncertainties \cite{FW10}. Incorporating such modules into the estimation layer of the quadruple-tank system could significantly enhance its resilience and reliability.

A final frontier lies in \textit{control-oriented learning}, which aims to bridge model-based control with data-driven modeling. Recent research has explored robust predictor-based architectures that provide certified guarantees in the presence of learning uncertainty \cite{FW11}, online adaptation to changing environments through real-time learning \cite{FW12,FW13}, and data-efficient identification strategies leveraging active exploration \cite{FW14}. In large-scale nonlinear systems such as the quadruple-tank, where unmodeled dynamics or unmeasurable disturbances are common, learning-based methods can complement classical design by improving prediction accuracy and enabling safer control decisions. A particularly promising direction is to fuse learning modules into distributed observers or controllers, leading to hybrid systems capable of adapting to both parametric uncertainty and structural changes over time \cite{FW15}.

Overall, these extensions promise to enhance the adaptability, robustness, and scalability of distributed control systems in real-world applications.

\bibliographystyle{ieeetr}    
\bibliography{reference}

\end{document}